\documentclass[aip,graphicx,reprint,amsmath,amssymb,]{revtex4-1}
\usepackage[dvips]{graphicx}
\usepackage{comment}
\usepackage{amsmath}
\usepackage{tabularx}

\draft 

\begin{document}


\title{Evidence for topological surface states in metallic single crystals of Bi$_{2}$Te$_{3}$} 



\author{Sourabh Barua}
\email[]{sbarua@iitk.ac.in}
\affiliation{Department of Physics, Indian Institute of Technology Kanpur}


\author{K. P. Rajeev}
\affiliation{Department of Physics, Indian Institute of Technology Kanpur}

\author{Anjan K. Gupta}
\affiliation{Department of Physics, Indian Institute of Technology Kanpur}

\date{\today}

\begin{abstract}
Bi$_2$Te$_3$ is a member of a new class of materials known as topological insulators which are supposed to be insulating in the bulk and conducting on the surface. However experimental verification of the surface states has been difficult in electrical transport measurements due to a conducting bulk. We report low temperature magnetotransport measurements on single crystal samples of Bi$_2$Te$_3$. We observe metallic character in our samples and large and linear magnetoresistance from 1.5 K to 290 K with prominent Shubnikov-de Haas (SdH) oscillations whose traces persist upto 20 K. Even though our samples are metallic we are able to obtain a Berry phase close to the value of $\pi$ expected for Dirac fermions of the topological surface states. This indicates that we might have obtained evidence for the topological surface states in metallic single crystals of Bi$_2$Te$_3$. Other physical quantities obtained from the analysis of the SdH oscillations are also in close agreement with those reported for the topological surface states. The linear magnetoresistance observed in our sample, which is considered as a signature of the Dirac fermions of the surface states, lends further credence to the existence of topological surface states.
\end{abstract}

\pacs{73.25.+i, 72.20.My, 73.20.At, 71.70.Di}

\maketitle 

\section{Introduction}
Topological insulators constitute a new phase in condensed matter physics and have attracted considerable attention due
to the interesting physics underlying them and the potential applications they promise. The topological insulators are
different from ordinary insulators in the topology of their band structure. The band structure of the topological
insulators have states confined to the surface which span the band gap of the bulk band structure and as a result the
surface of a topological insulator is conducting despite having an insulating bulk.\cite{Hasan,Qi} The interesting physics does not end here; these surface states are protected from opening of a gap at the $\Gamma$ point in the Brillouin zone due to Kramer's degeneracy as long as time reversal symmetry is not broken and the dispersion is linear near this point. Moreover the spin is locked to the direction of momentum in these states and as a result backscattering is prohibited unless the impurities are magnetic, which can cause a spin flip. Bi$_{1-x}$Sb$_{x}$, Bi$_{2}$Se$_{3}$, Bi$_{2}$Te$_{3}$, Sb$_{2}$Te$_{3}$ and Bi$_{2-x}$Sb$_x$Te$_{3-y}$Se$_y$ (BSTS) are the main topological insulators that have been discovered. However most of the reported data on these materials has a conducting bulk due to defects and vacancies and as a result the detection of the surface states in transport measurements has been rather difficult.\cite{Checkelsky,Eto,Butch,Cao,Ren,Xia,Ren2,Analytis,Barua} A way to circumvent this problem has been to study Shubnikov-de Haas (SdH) oscillations in the magnetoresistance of these samples and to see their dependence on the tilted magnetic field to single out oscillations due to two-dimensional (2D) Fermi surfaces corresponding to the surface states. The magnitude of the oscillations in case of a 2D Fermi surface are expected to be dependent on the perpendicular component of the magnetic field.\cite{Ren,Analytis} A feature of the magnetoresistance of these topological insulators that stands out is its large magnitude that has been observed in a host of reports \cite{Taskin1, Qu, Analytis2, Bansal,  Butch, Checkelsky, Cao, Taskin2, Eto} and has come to the forefront when Tang et al.,\cite{Tang} Wang et al., \cite{Wang} and He et al.\cite{He} emphasized it in their publications. It is also observed that the magnetoresistance in addition to being large is also linear and non-saturating.\cite{Tang,Wang,He,Yue,Hamdou,He2} Such large and linear magnetoresistance, which does not saturate even at high fields, was first observed in polycrystalline Bismuth and other metals by Kapitza \cite{Kapitza, Kapitza2} and this was explained by Lifshits \cite{Lifshits} et al as due to open Fermi surfaces of the metals. However Bismuth has a small and closed Fermi surface. Abrikosov then formulated a theory in which linear magnetoresistance would be possible at high fields, but he considered the attainment of such high fields as being physically impossible.\cite{Abrikosov} However, when Yang et al.\cite{Yang} rediscovered the linear magnetoresistance in Bismuth, this led to a connection being established between Abrikosov's theory and linear magnetoresistance.\cite{Abrikosov} Large magnetoresistance observed in non-magnetic silver chalcogenides by Xu et al. \cite{Xu} was explained by Abrikosov \cite{Abrikosov2} using his quantum linear magnetoresistance theory. Then Hu and Rosenbaum \cite{Hu} showed that it was possible to obtain linear magnetoresistance in InSb due to both quantum as well as classical reasons. Quantum linear magnetoresistance was reported in multilayer graphene by Friedman et al.\cite{Friedman} Qu et al. \cite{Qu} reported linear magnetoresistance in single crystals of Bi$_{2}$Te$_{3}$. However they reported magnetoresistance of both metallic and non-metallic samples. The linear magnetoresistance was observed only in case of the insulating crystals. The metallic crystals showed a quadratic
dependence on magnetic field at low fields. We report large, linear and non-saturating magnetoresistance in metallic
crystals of Bi$_{2}$Te$_{3}$. Tang et al. \cite{Tang} reported linear magnetoresistance in nanoribbons of topological
insulator Bi$_{2}$Se$_{3}$ and later He et al. \cite{He} reported similar linear magnetoresistance in Bi$_{2}$Se$_{3}$
thin films.

\begin{figure}[h!]
\includegraphics[width = 0.5\textwidth]{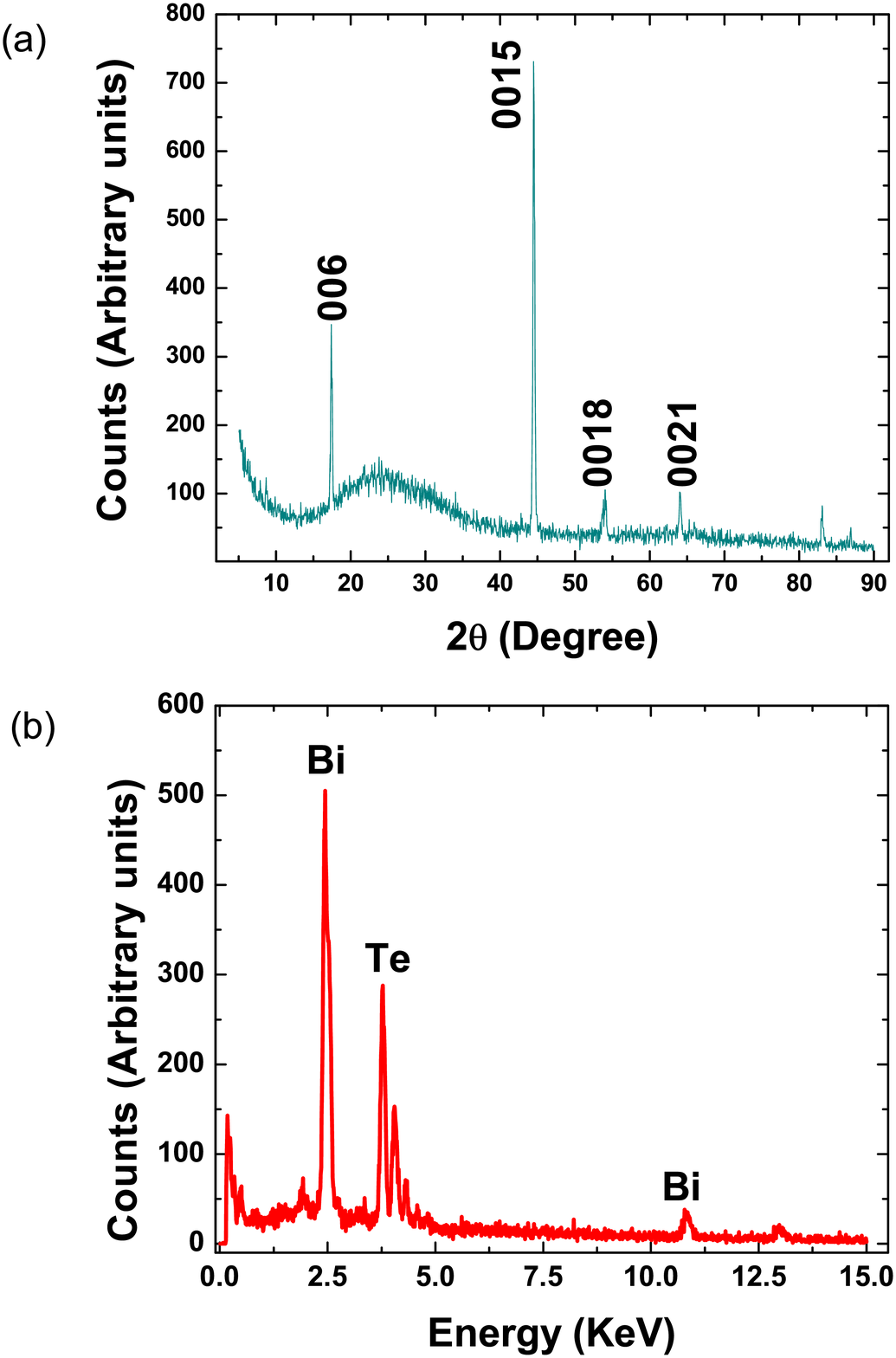}
\caption{(a) X-Ray diffraction pattern for a cleaved sample from the large Bi$_{2}$Te$_{3}$ single crystal. (b) EDX spectrum of another cleaved sample from the same Bi$_{2}$Te$_{3}$ single crystal. }  \label{Fig1}
\end{figure}
\begin{figure}[h!]
\includegraphics[width = 0.5\textwidth]{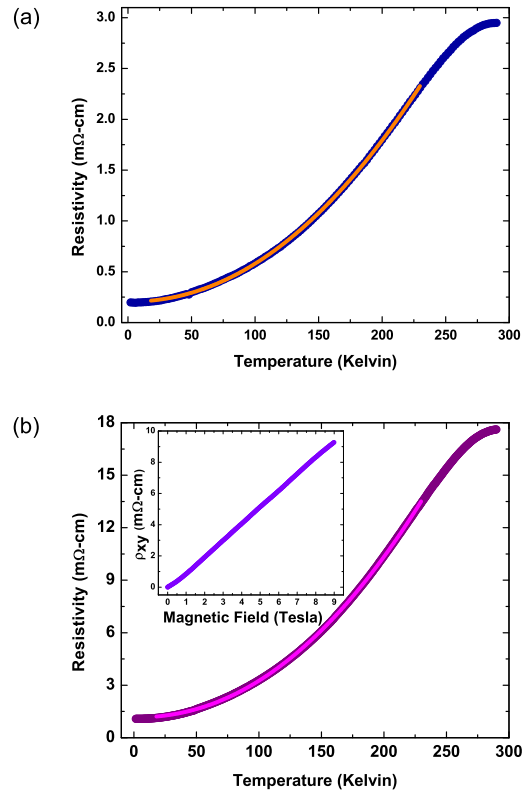}
\caption{(a) Resistivity versus temperature for cleaved single crystal of Bi$_{2}$Te$_{3}$ (Sample 1). The solid orange curve is a fit in the temperature range 18-230 K which yields a T$^{2.085}$ dependence of resistivity on temperature. (b) Resistivity versus temperature for cleaved single crystal of Bi$_{2}$Te$_{3}$ (Sample 2). The solid pink curve is a fit in the temperature range 18-230 K which yields a T$^{2.129}$ dependence of resistivity on temperature. Inset: Hall resistivity against magnetic field at 1.5 K for the same sample.}  \label{Fig2}
\end{figure}

\section{Experimental Details}
We performed electrical transport measurements on Bi$_{2}$Te$_{3}$ samples obtained from a natural single crystal. Thin samples were cleaved from this large single crystal using a razor blade and further thinned down by peeling off successive layers with scotch tape. Since Bi$_{2}$Te$_{3}$ has a layered crystal structure, it cleaves easily perpendicular to the c-axis with the cleaved single crystals having their flat surfaces perpendicular to the c-axis. In figure \ref{Fig1}(a), we show an X-ray diffraction pattern obtained for one of the cleaved single crystals using a PANalytical-X'Pert PRO diffractometer with Cu K$\alpha$ radiation and in figure \ref{Fig1}(b) we show the energy dispersive X-ray (EDX) spectrum for our sample which gives the chemical composition of our sample as 41.5 $\pm$ 0.1 \% Bi and 58.5 $\pm$ 0.1 \% Te by atomic percentage. The cleaved crystals were shaped in the form of rectangular bars, and the resistivity measurements were done using standard four-probe technique. The Hall measurement data reported for some of the samples was done simultaneously along with the resistivity measurement. The resistivity versus temperature, magnetoresistivity measurements and Hall measurements were all performed in an ICEoxford make closed cycle refrigerator $^{DRY}$ICE$^{VTI}$ which has a temperature range from 1.3 K to 300 K and provides a magnetic field up to 9 Tesla.

\begin{figure}[b!]
\includegraphics[width = 0.5\textwidth]{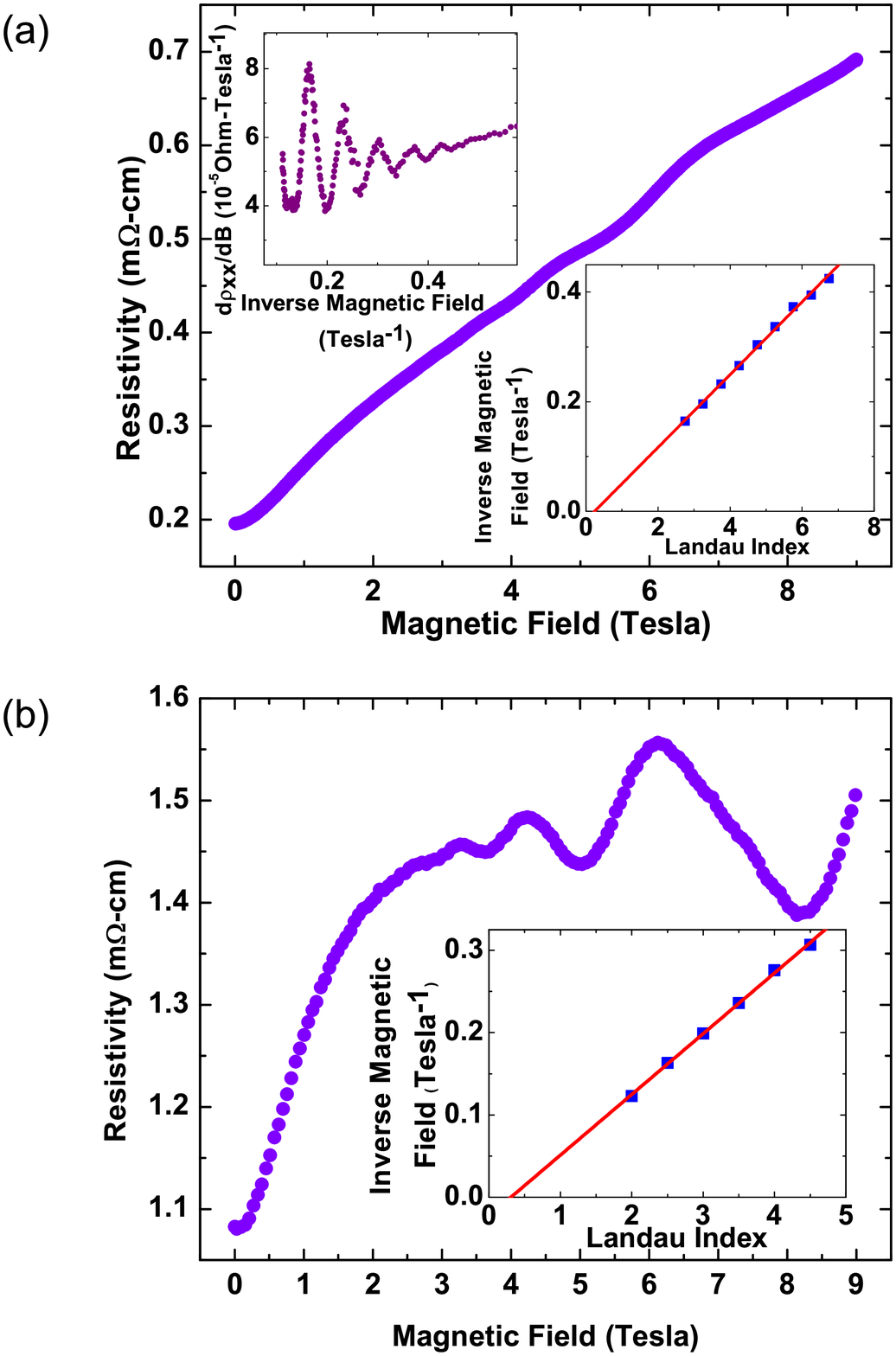}
\caption{(a) Magnetoresistance at 1.5 K for sample 1. Top inset:$\frac{d\rho_{xx}}{dB}$ plotted against inverse magnetic field at 1.5 K. Bottom inset: Landau level index plot for the oscillations shown in top inset with the solid squares representing the maxima and minima in the derivative of resistivity with respect to magnetic field. (b) Magnetoresistance at 1.5 K for sample 2. Inset: Landau level index plot for the oscillations with the solid squares representing the maxima and minima in the magnetoresistivity.} \label{Fig3}
\end{figure}

\section{Results}
The results of the resistivity versus temperature measurements for two different single crystal samples cleaved from the same bulk single crystal of Bi$_{2}$Te$_{3}$ are shown in figure \ref{Fig2}. The temperature dependence of the resistivity show that our samples are metallic. The residual resistivity ratio (RRR) in case of the first sample (sample 1) is about 14 and for the second sample (sample 2) is about 16 and these lie within the range reported in literature for this material\cite{Bos,Hor} with a higher RRR generally signifying greater crystallinity.\cite{Koyano} We also fit the resistivity data for both sample 1 and sample 2 to the equation $\rho = a + bT^n$ in the temperature range 18-230 K and we obtain values of $n$ = 2.085 $\pm$ 0.005 and 2.129 $\pm$ 0.001 respectively, which suggests fermi liquid behaviour. The uncertainty in the power of temperature $n$ was estimated by using the bootstrap method of finding the estimates of uncertainty in the parameters of a fit.\cite{Press} In the inset of figure \ref{Fig2}(b), we show the results of Hall measurement performed on sample 2. The results show that our single crystal is p-type in nature and a linear fit to the Hall voltage versus magnetic field above 1 Tesla yields a bulk charge carrier concentration $n_b^{Hall}$ = 5.87$\times$$10^{17}$ cm$^{-3}$. Bi$_{2}$Te$_{3}$ crystals are known to be p-type due to anti-site defects of Bi on Te sub-lattice sites.\cite{Hor,Hashibon,Rischau}

In figure \ref{Fig3} we show the magnetoresistance  at 1.5 K for samples 1 and 2, whose temperature variation of resistivity was shown in figure \ref{Fig2}. The resistivity shows clear oscillations at high magnetic fields in both cases. However in case of sample 2 the oscillations are more prominent and the background magnetoresistivity deviates from a linear dependence and saturates at high fields. These kind of oscillations in the magnetoresistance are commonly seen in metals and are known as the SdH oscillations \cite{Kittel}. SdH oscillations have been reported in this system earlier.\cite{Qu,Rischau,Veldhorst,Hamdou} One can extract the size of the Fermi surface from the SdH oscillations and also its shape from the dependence of the oscillations on the angle between the magnetic field and different crystal axes using the semiclassical quantization relation
\begin{equation} \label{Eqn1}
2\pi(n + \gamma) = \pi k_F^{2} \frac{\hbar}{eB}
\end{equation}
where $\hbar$ is planck's constant, $e$ is electron charge, $n$ is the landau level index, $B$ is the magnetic field, $\pi k_F^2$ is the cross-sectional area at the extremum of the Fermi surface and $\gamma$ is a phase correction which has the value 1/2 for normal fermions  with parabolic energy dispersion and 0 for Dirac fermions with linear energy dispersion.\cite{Qu,Kittel,Fuchs,Wright,Taskin4}

Since for sample 1 the oscillations are not very prominent we took the derivative of resistivity with respect to magnetic field and plotted it against the inverse of magnetic field as shown in the inset of figure \ref{Fig3}(a). Clear SdH oscillations are visible at fields above 2 T in the derivative of resistivity . The minima in resistivity correspond to integer values $n$ and the maxima to $n + 1/2$.\cite{Qu} So minima in derivative of resistivity with magnetic field will correspond to $n + 1/4$ and maxima to $n + 3/4$.\cite{Ren} A plot of the inverse field values at which maxima and minima in oscillations occur against the integer values known as the Landau indices which are used to label these peaks, will be a straight line and is known as a Landau level (LL) fan diagram. Such a LL fan diagram for sample 1 is shown in the bottom inset of figure \ref{Fig3}(a) where we have plotted the inverse field values at which maxima and minima in derivative of resistivity occur against the Landau indices after labeling the peaks  and valleys with the proper $n +3/4$ and $n+1/4$ values. The slope of this straight line, which is also equal to the frequency $f$ of these oscillations, is given by
\begin{equation} \label{Eqn2}
\frac{1}{f} = \frac{2\pi e}{\hbar} \frac{1}{\pi K_f^2}
\end{equation}
Upon performing a linear fitting of the LL fan diagram in figure \ref{Fig3}(a) we obtain a slope of 0.066 and the value obtained for intercept on the $n$ axis is $ 0.242$. The SdH oscillations can give further information about the Berry phase of the charge carriers and the topological surface state charge carriers are supposed to have a Berry phase of $\pi$ or a Berry phase factor of 1/2 because of their Dirac dispersion relation.\cite{Ren} We shall comment further on the extraction of the Berry phase from the LL fan diagram in the section in the discussion on Berry phase. Also from the slope of the fit to the LL fan diagram the Fermi wave vector $k_F$ turns out to be 0.021 \AA$^{-1}$. If we assume that this corresponds to a two-dimensional (2D) Fermi surface (FS), then the surface carrier concentration\cite{Ren} is $n_{s} = k_F^2/4\pi = 3.5 \times 10^{11}$ cm$^{-2}$ while, on the other hand, if we assume a three-dimensional (3D) FS, we get the bulk carrier concentration\cite{Kittel} as $n_{b}^{SdH} = k_F^3/3\pi^{2} = 3.1\times10^{17}$ cm$^{-3}$.

In case of sample 2, as the oscillations were much more pronounced, it was possible to use the maxima and minima in magnetoresistivity itself to index the Landau levels with the maxima labelled with proper $n + 1/2$ values and minima with $n$ values. In the bottom inset of figure \ref{Fig3}(b) we show the Landau level plot for sample 2. From the slope of the fit we obtain a $k_F$ value of 0.0203 \AA$^{-1}$ and the intercept on $n$ axis is $0.308$. From the $k_F$ value obtained in case of sample 2 we get surface carrier concentration of $n_{s}$ = 3.28$\times10^{11}$ cm$^{-2}$ if we assume a 2D FS and a bulk carrier concentration of $n_{b}^{SdH}$ = 2.83$\times10^{17}$ cm$^{-3}$ if we assume a 3D FS. If we compare this with the Hall concentration obtained for sample 2 then we see that the SdH carrier concentration is smaller by a factor of 2.07. In the work done by Rischau et al. the carrier concentration obtained from SdH oscillations had to be multiplied by a factor of 6 to obtain the total carrier concentration and this was accounted for by taking into consideration the six valley model for the lowest valence band and highest conduction band of Bi$_{2}$Te$_{3}$.\cite{Rischau} Kulbachinskii et al. also reported that the carrier concentration obtained from SdH was less than that obtained from Hall effect and this was explained as being due to the filling of a second lower valence band.\cite{Kulbachinskii} We could not make a similar comparison for sample 1 as we had not performed Hall measurement for that sample. A third sample for which we made both Hall and resistivity measurements gave values of Hall concentration of 1.74 $\times$ 10$^{18}$ cm$^{-3}$ and from its Landau level index plot we got a SdH carrier concentration of 3.1 $\times$ 10$^{17}$ cm$^{-3}$ which is smaller by a factor of 5.3. This would suggest that the SdH oscillations in our sample are originating from the bulk or it could also be the case that the SdH oscillations have a surface origin as is discussed in section on mobility in the discussion. The variation in Hall concentration in different samples cleaved from the same bulk single crystal has been previously seen in Bi$_2$Te$_3$ and Bi$_2$Te$_2$Se single crystals also.\cite{Qu,Jia}

\begin{figure}[h!]
\includegraphics[width = 0.5\textwidth]{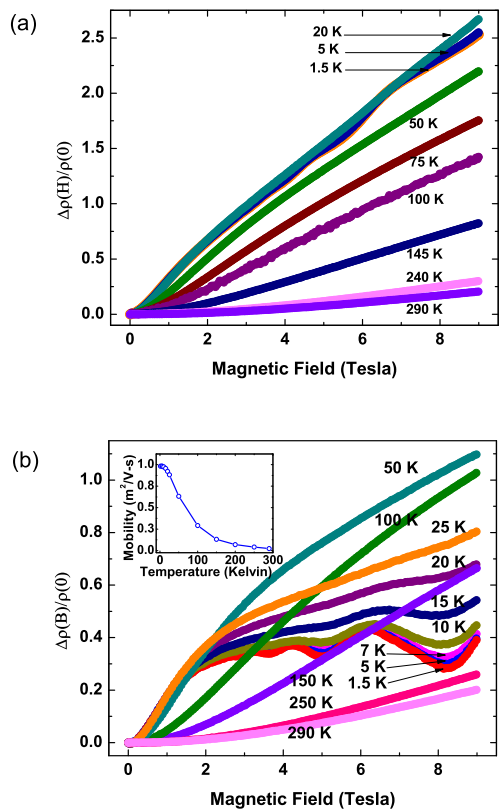}
\caption{(a) Relative change in resistivity with magnetic field at various temperatures for sample 1. The magnetoresistivity shown for 240 K and 290 K are from a different sample cleaved from the same single crystal. (b) Relative change in resistivity with magnetic field at various temperatures for sample 2. The inset shows the temperature variation of Hall mobility for sample 2  } \label{Fig4}
\end{figure}

The magnetoresistance at various temperatures is plotted in figure \ref{Fig4}. From the magnetoresistance of sample 1 in figure \ref{Fig4}(a) we note two significant observations. Firstly there is a large change in magnitude of the resistivity with magnetic field which is as high as 267\% at low temperatures upto 25 K and 20\% at 290 K. Such large magnetoresistance is seen in very few materials and has great potential for applications. The second significant observation is that the resistivity increases linearly with the magnetic field after an initial parabolic increase and also does not saturate at high fields. In case of sample 2, as seen in figure \ref{Fig4}(b), at low temperatures where the SdH oscillations are very prominent, the magnetoresistance tends to saturate. However the linear nature returns as temperature rises and the maximum magnetoresistance is obtained at intermediate temperatures with the magnitude of magnetoresistance decreasing as temperature rises to room temperature.

\section{Discussion}

The study of the SdH oscillations can give further information regarding the mobility, Berry's phase and the effective
mass of the charge carriers. To study the SdH oscillations, first a linear background was subtracted
from the magnetoresistance data in case of sample 1 and a parabolic background in case of sample 2. The SdH oscillations after such subtraction at different temperatures is shown in figure \ref{Fig5}. One standard expression for the SdH oscillation is
\begin{equation}
\label{Eqn3}
\frac{\Delta \rho}{\rho_0} = \frac{5}{2}\sum \limits_{r = 1}^{\infty} b_r cos\left(\frac{2\pi E_F}{\hbar \omega_c}r - \frac{\pi}{4}\right)
\end{equation}
with the amplitude $b_r$ given by
\begin{equation*}
\begin{split}
b_r = \frac{{(-1)}^r}{\sqrt{r}} \left({\frac{\hbar \omega_c}{2E_F}}\right)^{\frac{1}{2}} \frac{2\pi^2rk_BT/\hbar \omega_c}{sinh(2\pi^2rk_BT/\hbar \omega_c)} \\ \times exp\left(-\frac{2\pi^2rk_BT}{\hbar \omega_c}\right) cos\left(\frac{\pi r g m^*}{2 m_e}\right)
\end{split}
\end{equation*}
$k_B$, $g$, $m^*$, $\omega_c = eB/m^*$, and $T$ are Boltzmann constant, Land\'e g factor, cyclotron frequency and temperature respectively.\cite{Schneider} $T_D$ is the Dingle temperature and takes into account the broadening of the Landau levels and is related to the lifetime $\tau$ of a state by the relation $ T_D = \hbar/ 2\pi k_B \tau$. A similar expression which has been used for SdH oscillations in the case of topological insulators and explicitly includes the berry phase factor $\beta$ is\cite{Tang}
\begin{equation}
\label{Eqn4}
\Delta R(B) = A exp(-\pi/\mu B) cos[2\pi(B_F/B +0.5 + \beta)]
\end{equation}
with the amplitude $A$ having the following temperature dependence
\begin{equation*}
A \propto \frac{2\pi^2 k_B T/\hbar\omega_c}{sinh(2\pi^2 k_B T/\hbar\omega_c)}
\end{equation*}

We used the following combined form of equations \ref{Eqn3} and \ref{Eqn4} to fit our SdH oscillation data
\begin{equation}
\label{Eqn5}
\begin{split}
\Delta\rho = A\exp(-\pi/(\mu B))B^{1/2}\cos[2\pi(B_{F}/B+0.5+\beta)] \\ + \alpha -\gamma B + \delta B^2
\end{split}
\end{equation}
where $B$ is the magnetic field, $A$ is the amplitude, $B_{F}$ is the frequency of oscillation which are periodic in inverse of magnetic field, $\mu$ is mobility of the charge carrier and $2\pi\beta$ is the Berry's phase. We found out that the inclusion of the $B^{1/2}$ term into our expression which was missing in equation \ref{Eqn4} and is implicit in equation \ref{Eqn3} through the $\omega_c^{1/2}$ term was necessary in order to obtain a good fit. The parabolic contribution , $\alpha - \gamma B + \delta B^2$, takes care of any remnant background in the oscillations and was found to improve the quality of the fit. This is evident from the fact that the coefficient of determination ($R^2$) for the fit with parabolic contribution is 0.9811 while that for the fit without the parabolic contribution is 0.8635.

We fitted the oscillations in resistivity at 1.5 K to this equation and the resulting fit alongwith the original data are shown in figure \ref{Fig6}. The values of $A$ , $\mu$, $\beta$ and $B_{F}$ thus obtained for both samples along with the uncertainty in these parameters obtained by the bootstrap method of finding estimates of uncertainty in the parameters of a fit are given in the caption of figure \ref{Fig6}.\cite{Press} 

\subsection{Magnitude of SdH oscillations}

As pointed out earlier in the results section, the amplitude of SdH oscillations in sample 2 is much larger than that of sample 1. This can be explained on the basis of equation \ref{Eqn3} from which we see that the magnitude of oscillations is proportional to the background resistivity $\rho_0$. Thus it will be higher in case of sample 2 whose resistivity at zero field is of the order of 1 m$\Omega$-cm and this is about 5 times higher than that of sample 1 whose resistivity at zero field is of the order 0.2 m$\Omega$-cm. In fact the amplitude of oscillations at a particular field and temperature is about five times higher for sample 2 as compared to sample 1. That the magnitude of oscillation is proportional to the zero field resistivity at particular temperature is verified by the fact that the ratio of  magnitude of oscillations to zero field resistivity $\Delta \rho / \rho_0$ for sample 1 and sample 2, which are 0.071 and 0.067 respectively, are of similar magnitude. We make a similar observation in the work of Schneider et al. \cite{Schneider} who point out that amplitude of SdH oscillations decreases across samples with increasing electron concentration but we also note that resistivity also decreases across those samples. Also Xiong et al. \cite{Xiong} mention that although SdH oscillations were observed in all of their samples of Bi$_2$Te$_2$Se, they were the largest for only those samples whose resistivity exceeded 4 Ohm-cm.

\subsection{Berry phase}
There are two ways of extracting the Berry phase of the charge carriers from the SdH oscillations; one is from the LL fan diagram and other is by fitting the SdH oscillations to equation \ref{Eqn5}. However, we have noticed in the literature, an inconsistency in obtaining the Berry phase from the LL fan diagram using the semiclassical quantization relation given by equation \ref{Eqn1}. Some authors report that the phase correction $\gamma$ in equation \ref{Eqn1} is related to the Berry phase factor $\beta$ as $\gamma = \frac{1}{2}- \beta$ and that $\gamma$ = 1/2 ($\beta$ = 0) for normal fermions and 0 ($\beta$ = 1/2) for Dirac fermions.\cite{Qu,Wright,Taskin4} If this is correct than from equation \ref{Eqn1} we see that the intercept on $n$ axis is equal to $- \gamma$ and not $\beta$. However in most experimental papers, it is assumed that intercept on the $n$ axis is equal to $\beta$ itself.\cite{Xiong,Hamdou,Bao,Taskin2,Ren,Ren3,Tian,Taskin3,Yan} Furthermore, the same expression given in equation \ref{Eqn1}, is interpreted in a contrasting manner, in the sense that $\gamma$ in equation \ref{Eqn1} is now taken as actually $\beta$, the Berry phase factor.\cite{Hamdou,Tian} In fact, different forms of equation \ref{Eqn1}, like $ A_N = 2\pi e/\hbar B (N_{min} +\gamma -1/2)$ and $N = nh/e (1/B) - 1/2$ are also reported and which give the intercept on $n$ axis as $\beta$.\cite{Yan,Sacepe} The latter equation is specifically meant for the case of massless Dirac fermions.

Coming to the issue of determining the Berry phase of our samples from the LL fan diagram, we see that if we assume that the intercept on $n$ axis gives the value of $- \gamma$ then we have $- 0.242$ and $- 0.308$ as the values of $\gamma$ for samples 1 and 2 respectively. Since $\gamma$ is related to the Berry phase $\beta$ as $\gamma = \frac{1}{2}- \beta$ we have $\beta$ of 0.742 for sample 1 and 0.808 for sample 2. The values of $\beta$ that we had obtained from the fit to SdH oscillations at 1.5 K are (0.402 $\pm$ 0.009) and (0.359 $\pm$ 0.007) for the two samples 1 and 2 respectively and these are nearer to 0.5 which suggests a possibility that the oscillations are from Dirac charge carriers. However, if we assume that the intercept on $n$ axis gives the value of $\beta$, as is the convention in most experimental papers, then we see that the values of $\beta$ from the LL fan diagram are simply $0.242$ and $0.308$ for samples 1 and 2 respectively and in that case they are closer to the values obtained from the SdH fit.

\subsection{Mobility of charge carriers}
From the fit of SdH oscillations at 1.5 K we obtain a mobility of (2210 $\pm$ 40) cm$^2$/V-s and (2800 $\pm$ 100) cm$^2$/V-s for sample 1 and 2 respectively. The Hall mobility at 1.5 K for sample 2 from the Hall concentration and resistivity data turns out to be 9833 cm$^2$/V-s. This shows a striking resemblance to the situation in case of Bi$_2$Se$_3$ nanoplates where the Hall mobility of 8800 cm$^2$/V-s was higher than the SdH mobility of 1300 cm$^2$/V-s for the in-plane field and these SdH oscillations with the in-plane field were attributed to the topological surface states of the 2D sidewalls.\cite{Yan2} Furthermore, like in sample 2, in this sample also, the Hall concentration of 5.2 $\times$ 10$^{18}$ cm$^{-3}$ is about 2.8 times higher than the SdH concentration of 1.8 $\times$ 10$^{18}$ cm$^{-3}$ which can be obtained from the reported fermi wave vector value assuming a 3D FS. Thus the different Hall and SdH concentrations along with the different Hall and SdH mobility in our samples could imply that the SdH oscillations originate from the surface states, especially since the Berry phase extracted from the SdH oscillations point to presence of Dirac charge carriers which are supposed to be in the surface states. However it should be noted that, the SdH mobility can be smaller than the Hall mobility even if both have a bulk origin, because in the former the quantum lifetime $\tau$ plays an important role whereas in the latter the transport lifetime $\tau_{tr}$ is the important physical quantity and generally $\tau_{tr} > \tau$. This is due to reason that $1/\tau_{tr}$ acquires a factor of (1- cos$\phi$) due to spatial averaging, where $\phi$ is the scattering angle, while $1/\tau$ does not and at low temperatures small angle scattering can dominate, which will make the factor (1-cos$\phi$) small and in turn make $\tau_{tr}$ larger.\cite{Eto} Nevertheless, this necessarily does not rule out the possibility that the SdH oscillations have a surface origin as even in topological surface states, the small angle scattering will dominate since backscattering is forbidden and hence $\tau_{tr}$ will be larger.\cite{Qu2}

\subsection{Effective mass of charge carriers from temperature dependence of SdH oscillation}

\begin{figure}
\includegraphics[width = 0.5\textwidth]{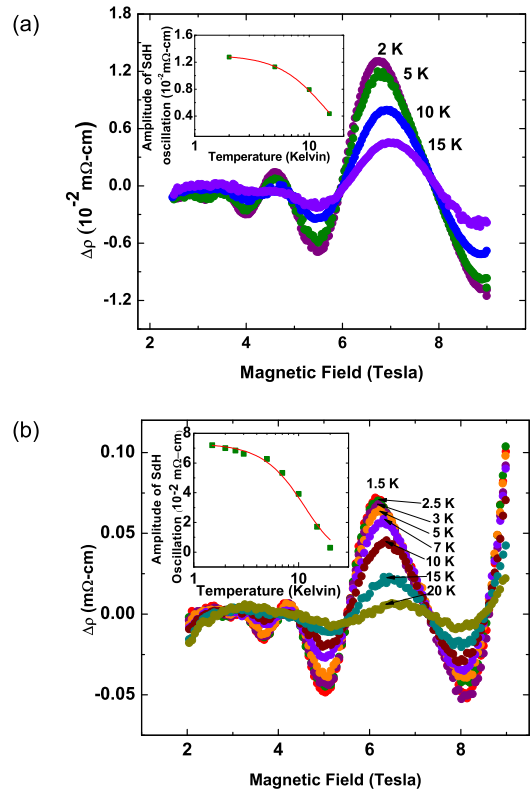}
\caption{(a) SdH oscillations at different temperatures after subtracting linear background for sample 1. Inset: Fit of the temperature dependence of amplitude of SdH oscillation in the magnetoresistivity at magnetic field of 6.87 T. (b) SdH oscillations at different temperatures after subtracting parabolic background in case of sample 2. Inset: Fit of the temperature dependence of amplitude of SdH oscillation in the magnetoresistivity at magnetic field of 5.028 T } \label{Fig5}
\end{figure}

\begin{figure}
\includegraphics[width = 0.5\textwidth]{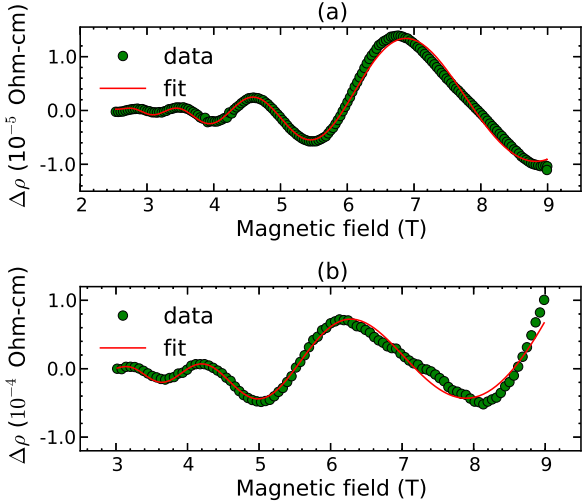}
\caption{Fit of the oscillation in magnetoresistivity at 1.5 K after subtracting background magnetoresistivity for sample 1(Panel(a)) and sample 2(Panel(b)). We obtained the values of (3.3 $\pm$ 0.2)${\times}$10$^{-5}$ Ohm-cm-Tesla$^{-1/2}$, (2210 $\pm$ 40) cm$^{2}$/V-s, (0.402 $\pm$ 0.009) and (14.06 $\pm$ 0.05) T for $A$, ${\mu}$, ${\beta}$ and $B_F$ respectively at 1.5 K for sample 1 and the values (1.6 $\pm$ 0.1)${\times}$10$^{-4}$ Ohm-cm-Tesla$^{-1/2}$, (2800 $\pm$ 100) cm$^{2}$/V-s, (0.359 $\pm$ 0.007), and (13.09 $\pm$ 0.04) T for A, ${\mu}$ ,${\beta}$ and $B_F$ respectively at 1.5 K for sample 2.}
\label{Fig6}
\end{figure}

The SdH oscillations get damped as temperature rises and as is evident from figure \ref{Fig5}(a) and (b), traces of the oscillations can be seen upto 15 K in sample 1 and upto 20 K in sample 2. The amplitudes of the SdH oscillation for a fixed value of the field were plotted for different temperatures for sample 1 and 2 in the inset of figure \ref{Fig5}(a) and (b) and they were fitted with the following expression.

\begin{equation}
\label{Eqn6}
\Delta\rho(T) = \Delta\rho(0)\frac{2 \pi^2 k_B T/ \hbar\omega_c}{\sinh(2 \pi^2 k_B T/ \hbar\omega_c)}
\end{equation}

We chose the maxima at 6.87 T for sample 1 and at 6.124 T for sample 2. From this fitting the extracted effective mass of the charge carrier turns out to be 0.09 $m_{e}$ for both samples, where $m_{e}$ is the free electron mass. Qu et al. \cite{Qu} reported a value of 0.1 $m_e$ for Bi$_{2}$Te$_{3}$.

\subsection{Fermi velocity, Fermi energy and electron mean free path}
 From the value of effective mass $m^*$ and Fermi wave vector, the Fermi velocity $v_F$  using $v_F = \frac{\hbar k_F}{m^*}$ comes out to be 2.7 $\times$ 10$^5$ m/s for sample 1 and 2.6 $\times$ 10$^5$ m/s for sample 2, which is close to the value of 1.4 $\times$ 10$^5$ m/s reported by Veldhorst et al. and 3.7 $\times$ 10$^5$ m/s reported by Qu et al. for Bi$_2$Te$_3$.\cite{Qu,Veldhorst} Similarly, the Fermi energy, $E_F$ comes out to 37 meV for sample 1 and 35 meV for sample 2 using a relativistic formula $E_F = m^* v_F^2$.\cite{Qu} Qu et al. reported $E_F$ in the range of 78-94 meV for the surface Dirac electrons in Bi$_2$Te$_3$. Comparing equations \ref{Eqn3} and \ref{Eqn4}, we see that the Fermi energy is related to the frequency of oscillation in inverse magnetic field as $E_F = (\hbar e /m^*)B_F$. Using this expression we obtain $E_{F}$ values of 18 meV and 16.8 meV for samples 1 and 2 respectively. The Fermi energy obtained from the frequency of SdH oscillations is almost half of that obtained from the Fermi velocity using the relativistic formula. We note that this was true in the case of a few other reports also, when we compared the Fermi energy calculated from the Fermi velocity and the frequency of SdH oscillations but the reason behind this is not clear.\cite{Hamdou,Yan2}

 The electron mean free path $l_e$ can be calculated from the SdH mobility and Fermi wave vector using the relation $\mu = e l_e/\hbar k_F$ using which we obtain $l_e$ of 31 nm and 37  nm for sample 1 and sample 2 respectively. The metallicity parameter $k_Fl_e$ then turns out to be 6.5 and 7.5 for samples 1 and 2 respectively. As comparison we note that $l_e$ values ranging from 105 nm to 219 nm \cite{Qu,Veldhorst,Hamdou,Tian} and $k_Fl_e$ values ranging from 13.8 to 66 have been reported for the surface states of Bi$_2$Te$_3$ in the literature.\cite{Qu,Hamdou,Tian}

\subsection{Saturation of magnetoresistance}
As pointed out in the results section, the magnetoresistance of sample 2 differs from that of the sample 1 at low temperatures. The magnetoresistance of sample 1 is large, linear and non-saturating with the magnitude dropping with rising temperature while in sample 2 the magnetoresistance at low temperatures has a tendency to saturate. The magnetoresistance however becomes linear at high fields once the temperature rises above 100 K and then onwards the magnitude of magnetoresistance decreases with rising temperature as in sample 1. This can be explained on the basis of the classical magnetoresistance theory according to which the condition for saturation to set in is that $\omega_c \tau \gg 1$ where $\omega_c$ is the Larmor frequency and $\tau$ is the collision time.\cite{Abrikosov} This condition can be re-written  as $\mu B \gg 1$ where $\mu$ is the mobility and B the magnetic field and from which we see that as mobility decreases the condition for saturation to set in is fulfilled at progressively higher values of the magnetic field. We have shown in the inset of figure \ref{Fig4}(b) the temperature dependence of Hall mobility for sample 2 from which it is evident that the change in mobility is very small till 25 K and then falls drastically as temperature rises. This coincides with the temperature range in which saturation is seen in the magnetoresistance and as temperature rises the mobility decreases leading to the increase in value of magnetic field at which saturation will set in thus making it unattainable within the maximum magnetic field obtainable in our experiments.

\subsection{Large and linear magnetoresistance}

The magnetoresistance of our Bi$_{2}$Te$_{3}$ samples is characterized by a large magnitude which is substantial even at room temperature and a linear dependence of the resistivity on the magnetic field after an initial parabolic rise that does not saturate even at a field of 9 T. The only exception is that the magnetoresistance at low temperatures in sample 2 shows a saturating behaviour. But even in this sample linear magnetoresistance makes a comeback at temperatures above 100 K  and thereafter it is similar to that of sample 1 in terms of linearity and magnitude. Such large and linear non-saturating magnetoresistance has been reported earlier in insulating Bi$_{2}$Te$_{3}$.\cite{Qu,Wang} Qu et al. reported magnetoresistance studies on metallic crystals of Bi$_{2}$Te$_{3}$ but the magnetoresistance was not linear.\cite{Qu} Recently, linear magnetoresistance has been reported in metallic Bi$_{2}$Te$_{3}$ by Yue et al. and Hamdou et al.\cite{Yue,Hamdou} It should be noted though that Yue et al. report metallic samples even though these were taken from the same bulk crystal used in the earlier work of Wang et al. who reported insulating samples.\cite{Yue,Wang} Thus our report is among one of the few instances of large, linear magnetoresistance in metallic Bi$_2$Te$_3$ bulk single crystals.

From figure \ref{Fig4} it can be seen, that the magnitude of the magnetoresistance at a field of 9 T  is 267 \% at low temperature and 20 \% at 290 K for sample 1 while for sample 2 it is maximum at 50 K with a magnitude of 109 \% and at 290 K it is 20 \%. To compare it with other reports we note that Qu et al. \cite{Qu} reported a similar magnitude of 250 \% for their metallic crystals and about 180 \% for their insulating crystals, while Wang et al. \cite{Wang} reported magnetoresistance up to 625 \%. The large and linear magnetoresistance generally decreases in magnitude as the temperature rises, and it is observed in our samples also.\cite{Tang,Qu,He}

The linear magnetoresistance in topological insulators has been attributed to the presence of topological surface states with a Dirac dispersion relationship.\cite{Tang,He,Wang} Classical theory of magnetoresistance accounts for the quadratic dependence of the resistivity on magnetic field at low fields followed by saturation at high fields seen in normal metals.\cite{Abrikosov,Tang} The linear and non saturating magnetoresistance observed in silver chalcogenides, \cite{Xu} InSb, \cite{Hu} multi-layer graphene, \cite{Friedman} and in topological insulators can be explained by the theory of quantum magnetoresistance put forward by Abrikosov.\cite{Abrikosov2} The theory predicts that in case of a gapless and linear dispersion relation, it is possible to have a linear and non-saturating magnetoresistance. This holds true in case of topological insulators which have a surface state with linear dispersion relation. According to this theory, in order to observe linear magnetoresistance the magnetic field should be so high that only the lowest landau level is populated and also the temperature should be not so high that the thermal energy exceeds the Landau level separation. In terms of the charge carrier concentration, $n_{e}$ and effective mass of the charge carriers, $m^{*}$ the conditions to be satisfied by the field $B$ and temperature $T$ are

\begin{equation}
n_e << (\frac{eB}{\hbar})^{3/2} \qquad T < v \sqrt{eB\hbar}/k_B\label{Eqn7}
\end{equation}
where $v$ is taken to be 10$^{+8}$ cm/s and $\hbar$ is the Planck's constant.

We now check whether the conditions given in equation \ref{Eqn7} for Abrikosov's theory of linear magnetoresistance to be applicable are fulfilled in our sample. We note from figure \ref{Fig4}(a) that in our samples there is a changeover from quadratic dependence on magnetic field at low fields to a linear dependence at higher fields at around 2 T. Such a crossover has been seen in almost all instances of linear magnetoresistance reported for topological insulators as well as other materials.\cite{Qu,Hu,Tang,He,Friedman} Thus beyond the crossover field of 2 T the first condition of equation \ref{Eqn7} must be satisfied and for linear magnetoresistance to be seen at a field of 2 T we obtain the condition that the carrier concentration should be less than 1.67 $\times$ 10$^{17}$ cm$^{-3}$ which is clearly exceeded by the carrier concentration of 5.87$\times$$10^{17}$ cm$^{-3}$ obtained from Hall measurement. However if we convert the surface carrier concentration $n_{s} = 3.5 \times 10^{11}$ cm$^{-2}$ obtained from SdH oscillations to bulk concentration using the thickness of the sample which is 136 $\mu$m, then we obtain a bulk concentration of 2.6 $\times$ 10$^{13}$ cm$^{-3}$ which lies well below the limit. Since the SdH oscillations have already shown signs of Dirac states, evidence that the linear magnetoresistance is originating from the same surface states would further strengthen the case since linear magnetoresistance is supposed to rise in case of gapless linear dispersion states. As far as the condition for only the lowest Landau level being filled is concerned, we can see from the Landau index plot of figure \ref{Fig3} that even at high fields more than one Landau level is filled. However Hu et al. \cite{Hu} have shown that linear magnetoresistance can set in at low magnetic fields even when more than one Landau level is populated.

The second part of equation \ref{Eqn7} gives a temperature of nearly 421 K for a field of 2 T. This agrees with the fact that we observe linear magnetoresistance till room temperature. These observations would suggest that the linear magnetoresistance in our sample satisfies the criteria for observing quantum magnetoresistance thereby signifying that it is a signature of the topological surface states. However in light of the recent finding by Yue et al. from their angular dependent magnetoresistance studies, that the linear magnetoresistance in Bi$_{2}$Te$_{3}$ could have its origin in the bulk states,\cite{Yue} one has to exercise caution in attributing the linear magnetoresistance to the surface states. Moreover sample inhomogeneity, which can also lead to linear magnetoresistance but was previously considered to be not valid in the context of topological insulators\cite{Tang,Wang} has recently been shown to be behind the linear magnetoresistance  in Bi$_{2}$Se$_{3}$.\cite{He2} Thus the issue of the linear magnetoresistance in topological insulators remains an open question and further studies are necessary. The extremely large value of the magnetoresistance and its linearity alongwith the fact that it persists upto room temperature opens up possibilities of this material finding use in applications.

\section{Conclusion}
In summary, we report large, linear and non-saturating magnetoresistance in metallic single crystals of Bi$_{2}$Te$_{3}$
with magnetoresistance as large as 267 \% at low temperature, which persists up to room temperature although its magnitude is reduced. This is among the very few cases of linear magnetoresistance being reported in metallic single crystal of Bi$_{2}$Te$_{3}$ till now. Also we observe prominent SdH oscillations upto 20 K and the values of Berry's phase extracted from the Landau plot and fit to the SdH oscillations, especially from the latter, are close to the expected value of $\pi$ expected for the Dirac fermions of the topological surface states and point to the existence of topological surface states in our sample. The values of mobility, surface carrier concentration, effective mass and Fermi velocity extracted from the SdH oscillations are close to that reported in literature for the topological surface states. The values of Fermi energy, electron mean free path are also comparable to what has been reported earlier for surface states of Bi$_{2}$Te$_{3}$. The conditions for Abrikosov's quantum magnetoresistance to be seen are fulfilled by the 2D carrier concentration obtained from the SdH oscillations and this may imply that the large and linear magnetoresistance originates from these 2D states giving further indications of a topological surface state in our samples. Together the SdH oscillations and linear magnetoresistance provide evidence for topological surface states in the metallic single crystal of Bi$_{2}$Te$_{3}$.



%
%

%


\bibliographystyle{apsrev4-1}

\bibliography{Bi2Te3_paper_arxiv}

\end{document}